\documentclass[11pt,twoside]{article}
\usepackage{asp2014}

\aspSuppressVolSlug
\resetcounters

\newcommand{\domm}[1]{\ifmmode #1\else$#1$\fi}

\newcommand{\Msun}{\domm{M_\odot}}

\newcommand{\Mch}{\domm{M_{\rm Ch}}}

\newcommand{\Ma}{\domm{M_{\rm accretor}}}
\newcommand{\Mtot}{\domm{M_{\rm total}}}
\newcommand{\arepo}{\textsc{arepo}}

\newcommand{\gcc}{g cm$^{-3}$}

\begin{document}

\title{The Physics and End-Products of Merging CO WD Binaries}

\author{Chenchong Zhu
\affil{Department of Astronomy \& Astrophysics, University of Toronto, Ontario, Canada \email{cczhu@astro.utoronto.ca}}}

\paperauthor{Chenchong Zhu}{cczhu@astro.utoronto.ca}{ORCID_Or_Blank}{University of Toronto}{Department of Astronomy \& Astrophysics}{Toronto}{Ontario}{M5S 3H4}{Canada}

\begin{abstract}
The merger of two carbon-oxygen white dwarfs has long been theorized to lead to a massive carbon-oxygen or oxygen-neon white dwarf, accretion-induced collapse to a neutron star, or a type Ia supernova.  Determining which mergers lead to a particular outcome requires hydrodynamic simulations of the merging process.  I give a brief overview of the current understanding of mergers and their end-products derived from simulations, and show how temperature, rather than density or mass, most strongly determines a merging binary's subsequent evolution.  I then describe recent simulations that show mergers generate strong magnetic fields that could help drive a merger remnant to ignition.
\end{abstract}


\section{Introduction}

Of order 1\% of all carbon-oxygen white dwarfs (CO WD) will merge with another carbon-oxygen white dwarf within a Hubble time of their birth as a result of gravitational radiation sapping their orbital angular momentum.  The merging process begins when the less massive ``donor'' WD overflows its Roche lobe and transfers mass to its companion, the ``accretor''.  For the range of plausible CO WD mass ratios, mass transfer is unstable \citep{dan+11}, and leads to tidal disruption of the donor into a stream of material that coalesces with the accretor.  The merged object, or ``merger remnant'', consists of a degeneracy-supported core embedded within a partly thermally-supported envelope and ringed by a thick disk.  While the merger is expected to be a gravitational wave (and possibly electromagnetic) source (eg. \citealt{loreig09}), it is the possibility that a merger could instigate a type Ia supernova (SN Ia; \citealt{webb84, ibent84}) that has driven merger research for the past few decades.

It was long-held (eg. \citealt{yoonpr07}) that CO WD mergers can only produce Ias if the remnant disk slowly accretes onto the core, at $\sim10^{-6}$ $M_\odot$ yr$^{-1}$ over $\sim 10^5$ yr, until the core mass approaches the Chandrasekhar mass, \Mch, and the core central density becomes high enough ($\sim3\times10^9$ \gcc) to ignite pycnonuclear fusion and start a nuclear runaway (see \citealt{howe11} and \citealt{hill+13} for recent reviews).  If the accretion rate were closer to the Eddington rate $\sim10^{-5}$ $M_\odot$ yr$^{-1}$, one-dimensional calculations (eg. \citealt{nomoi85}) show that off-center convective carbon burning starts, which propagates inward over $\sim10^4$ yr to turn the merger remnant into an oxygen-neon white dwarf (ONe WD).  If the ONe WD has $M \gtrsim \Mch$, electron captures onto high-density material eventually force it to collapse to a neutron star - an ``accretion induced collapse'' (AIC; \citealt{saion85}) - though see \cite{guticg05} for how unburnt carbon could drive an ONe WD to  explode instead of collapse.  If neither a nuclear runaway nor collapse is triggered, the merger remnant would remain as a massive, rapidly-rotating and likely highly-magnetized (see Sec. \ref{sec:mag}) CO/ONe WD.  Indeed, mergers could explain a substantial fraction of high-field magnetic WDs (HFMWDs), a population weighted toward massive WDs that are either isolated or in interacting binaries (\citealt{garcla12} and references therein).  They may also be related to massive and magnetic hot DQ WDs, which appear to have surprisingly old dynamical ages for their mass (Bart Dunlap's conference proceeding).

Both the notion of slow accretion and the core reaching \Mch\ to start nuclear fusion have recently been challenged.  \cite{vkercj10} and \cite{shen+12} note that the remnant should be prone to magnetic instabilities such as the magnetorotational instability and Tayler-Spurit dynamo, and a simple Shakura-Sunyaev $\alpha$ prescription implies a (magnetic) viscosity-driven remnant spin-down over hours rather than $10^5$ yr.  Moreover, earlier calculations neglected that the merger remnant is hot - the virial temperature for accreted material in a typical merger is $\sim4\times10^8$ K, close to the carbon ignition threshold.  This raises the possibility that, in lieu of high density, high remnant core temperature during or after the merger triggers a nuclear runaway.  \cite{vkercj10} proposes these runaways occur in many CO WD mergers with \Mtot\ \textit{below} \Mch\, and are responsible for the lion's share of SNe Ia\footnote{There are also arguments against $M \gtrsim \Mch$ CO WD explosions producing SNe Ia independent of merger physics \citep{vkercj10, howe11, hill+13}.}.

At present, the possible outcomes of a CO WD binary merger remain the creation of a massive CO WD, off-center carbon ignition leading to a massive ONe WD, accretion induced collapse, and an SN Ia.  It remains to be determined which regions of the merger parameter space will lead to each of these outcomes.  

\section{The Merger Parameter Space}

Since it is currently impossible to directly observe mergers, they are instead investigated through three-dimensional hydrodynamic simulations.  In the last decade, the CO WD merger parameter space has been well-studied with various smoothed-particle hydrodynamics codes (SPH; eg. \citealt{spri10}).

Simulations of the most massive CO WD mergers see pockets of material near the accretor surface become hot enough that they could detonate, destroying the binary during merger \citep{pakm+10, pakm+11, pakm+12, dan+12, moll+14}.  The minimum masses required are still disputed, and may be as low as $\Ma \gtrsim 0.9\Msun$, $\Mtot \gtrsim 1.6 \Msun$, or as high as $\Ma \gtrsim 1.05\Msun$, $\Mtot \gtrsim 2.1 \Msun$ \citep{dan+14,moll+14}.  Further complicating matters is the $10^{-2}$ \Msun\ helium atmosphere on CO WDs.  \cite{rask+12} and \cite{pakm+13} show that for $\Ma \gtrsim 1\Msun$ this helium may detonate during the merger, catalyzing a second detonation at the center of the accretor that destroys the binary.  All of these massive mergers have $\Mtot$ well-above $\Mch$, meaning that even if they do survive the merger itself, post-merger evolution will carry them toward their inevitable end as SNe Ia or AIC.

Lower-mass mergers experience negligible nuclear burning during the merger proper \citep{loreig09, rask+12}, and survive coalescence; their remnants are studied in detail by \cite{zhu+13} and \cite{dan+14}.  \citeauthor{zhu+13} find that the structure of the merger remnant depends critically on the mass difference between the donor and accretor.  Dissimilar-mass mergers, with $\Delta M \gtrsim 0.1\Msun$, see little mixing between the donor and accretor, and produce remnants with a cold, slowly-spinning core surrounded by a much hotter envelope.  Similar-mass mergers, with $\Delta M \lesssim 0.1\Msun$, see both WDs substantially mixed, resulting in a remnant core that is both relatively hot and partly supported by rotation.  \citeauthor{dan+14}'s data set also show a strong dependence on $\Delta M$, but they do not find any remnants with hot cores.  This is because \citeauthor{dan+14} use synchronized WD binaries that include tidal bulges, which merge less violently than \citeauthor{zhu+13}'s spherical, unsynchronized WDs.  Whether or not merging CO WDs are synchronized is still not well-understood (see both works for discussion).

Two groups, \cite{schw+12} and \cite{ji+13}, have run two-dimensional simulations of post-merger evolution.  Both find that outward angular momentum transport over the course of a few hours converts the disk into a thermally supported envelope, and compresses and heats the core through a loss of rotational support.  \cite{schw+12} use a 0.6 - 0.9 \Msun\ remnant, and find carbon ignition at the base of the envelope, leading to stable conversion of carbon to oxygen and neon.  \cite{ji+13} start with a 0.6 - 0.6 \Msun\ remnant and see ignition of carbon at the core's center.  Since this material is highly degenerate, a nuclear runaway that may lead to an explosion is expected to follow.  These studies show that if nuclear ignition occurs, it will be where the merger remnant was initially the hottest, making the centrally-peaked temperature structure of similar-mass mergers necessary for starting a runaway.

\citeauthor{zhu+13} make a rough estimate of post-merger compression and heating of the hottest point in the remnant.  They find that dissimilar-mass mergers with $\Ma \gtrsim0.8 \Msun$ produce remnants that eventually ignite stable carbon fusion to oxygen and neon; of these, systems with $\Mtot > \Mch$ may eventually undergo AIC (or explode).  Similar-mass mergers with $\Ma \gtrsim 0.6 \Msun$ produce remnants that ignite a core nuclear runaway.  Systems that fit neither category will simply produce massive CO WDs.  \cite{dan+14}'s remnant parameter space has no remnants with hot cores, and therefore all remnants that ignite carbon burning will do so off-center and produce either ONe WDs or neutron stars (or perhaps explosions).

\section{Full MHD Merger Simulations}
\label{sec:mag}

The merging process features regions of intense shear and vorticity, which could feed magnetic dynamo processes to greatly amplify any initial fields the binary possessed.  Magnetic field growth has, however, not been considered in any WD merger simulations to date\footnote{Though MHD SPH has been used to simulate a binary neutron star merger \citep{pricr06}.} due to the difficulty in including them in SPH \citep{spri10}.  \citealt{zhu+14} utilize the MHD moving mesh code \arepo\ \citep{spri10b,pakms13} to model a fully magnetized 0.625 - 0.65 \Msun\ CO WD merger.  The initial conditions used are nearly identical to the 0.625 - 0.65 \Msun\ binary in \cite{zhu+13}, except that each WD is given a dipole magnetic field with a surface value of $10^3$ G.

\begin{figure}
\centering
\includegraphics[angle=0,width=1.0\columnwidth]{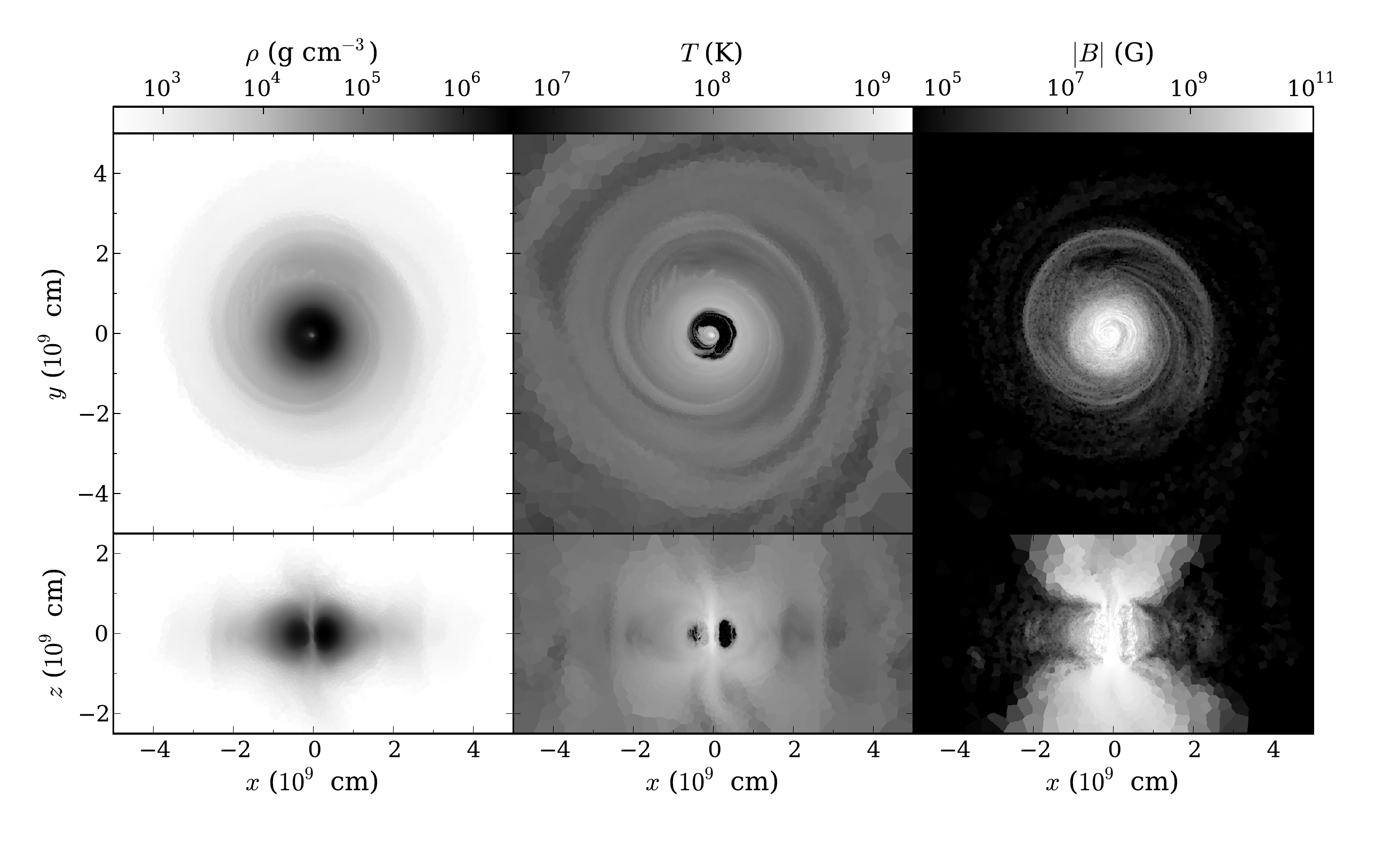}
\caption{Density, temperature and magnetic strength equatorial plane and polar profiles of a 0.625 - 0.65 \Msun\ CO WD merger remnant $\sim150$ s after coalescence.}
\label{fig:remfig}
\end{figure}

Fig. \ref{fig:remfig} shows the density, temperature and magnetic field profiles of the merger remnant.  During coalescence, shearing along the interface between the accretor and infalling donor material drives the exponential growth of the initial field, similar to what has been seen for neutron star merger simulations \citep{pricr06}.  This growth saturates following coalescence, leaving the remnant core with $\sim10^{11}$ G fields that extend out to the inner disk, as well as along the remnant rotation axis in a funnel structure.  This field configuration is qualitatively similar to the initial conditions used by \citeauthor{ji+13}, though our field has a significant toroidal component as well.

Unlike in SPH, the remnant core remains at low temperature despite the merger being similar-mass.  This is not due to the inclusion of magnetic fields, as they remain hydrodynamically weak until after coalescence, and is most likely because of more accurate gradient and hydrodynamic instability capture and a lower numerical viscosity in \arepo\ compared to SPH codes.  While this appears to rule out a hot core even for unsynchronized similar-mass mergers, the magnetic field becomes hydrodynamically important during post-merger evolution.  The field greatly increases the effective viscosity of the core and inner disk, converting differential rotation energy into thermal energy as a result.  In preliminary simulations, this heats the $3\times10^6$ \gcc\ degenerate center of the remnant core to $7\times10^8$ K, at the precipice of carbon ignition, within 1000 s after coalescence.  This could occur even if the merging binary were synchronized, making synchronization unimportant for whether a remnant can ignite a runaway - though \citeauthor{dan+14} find less mixing between donor and accretor in their similar-mass mergers compared to \citeauthor{zhu+13}, which could inhibit magnetic field growth.

The final fates of merging CO WDs is still far from fully understood.  For violent mergers, a parameter space study is needed to set constraints on which systems will experience either a helium or carbon detonation during the merging process.  For lower-mass mergers, the merger remnant space is well-sampled, but only a handful of simulations have been performed for post-merger viscous evolution \citep{schw+12, ji+13}.  The effects of strong magnetic fields must also be included - preliminary simulations over a range of masses indicate all merger remnants should have relatively strong magnetic fields, and similar-mass mergers have fields that permeate into their cores \citep{zhu+14}.  These fields would not only determine which remnants could ignite core nuclear runaways, but also imbue magnetar-strength magnetic fields to those remnants that undergo AIC instead of exploding.

Finally, simulations must be run of the long-term ($\sim10^4$ yr) thermal evolution of those remnants that never achieve fusion.  One-dimensional stellar evolution simulations are currently the only way to investigate evolution on these timescales, and they will be able to resolve photospheric properties of these objects.  Long-term magnetic field evolution could be charted in tandem (eg. Yevgeni Kissin's conference proceeding).  Such work will be able to provide a much-needed link between simulations and the population of observed massive WDs, allowing WD observations to act as a confirmation of the complex theoretical tools used to understand mergers.

\acknowledgements I thank Bart Dunlap and Yevgeni Kissin for insightful discussions on possible merger remnant observables and the long-term magnetic evolution of white dwarfs, respectively.  Thanks also to my supervisor Marten van Kerkwijk for his attention to detail when commenting on drafts of this work.

\bibliography{bibliography}  

\end{document}